# Sleep Apnea and Respiratory Anomaly Detection from a Wearable Band and Oxygen Saturation

Wolfgang Ganglberger, MSc*[1, 2], Abigail A. Bucklin, BA*[1], Ryan A. Tesh, BSc[1], Madalena Da Silva Cardoso[1], Haoqi Sun, PhD[1], Michael J. Leone, MSc[1], Luis Paixao, MD, MSc[1,3], Ezhil Panneerselvam, MD[1], Elissa M. Ye, MSc[1], B. Taylor Thompson, MD[4], Oluwaseun Akeju, MD[5], David Kuller, BSc[6†], Robert J. Thomas, MD[7†], M. Brandon Westover, MD, PhD[1†]

### Abstract

*Objective*: Sleep related respiratory abnormalities are typically detected using polysomnography. There is a need in general medicine and critical care for a more convenient method to automatically detect sleep apnea from a simple, easy-to-wear device. The objective is to automatically detect abnormal respiration and estimate the Apnea-Hypopnea-Index (AHI) with a wearable respiratory device, compared to an $SpO_2$ signal or polysomnography using a large (n = 412) dataset serving as ground truth.

*Methods*: Simultaneously recorded polysomnographic (PSG) and wearable respiratory effort data were used to train and evaluate models in a cross-validation fashion. Time domain and complexity features were extracted, important features were identified, and a random forest model employed to detect events and predict AHI. Four models were trained: one each using the respiratory features only, a feature from the $SpO_2$ (%)-signal only, and two additional models that use the respiratory features and the $SpO_2$ (%)-feature, one allowing a time lag of 30 seconds between the two signals.

*Results*: Event-based classification resulted in areas under the receiver operating characteristic curves of 0.94, 0.86, 0.82, and areas under the precision-recall curves of 0.48, 0.32, 0.51 for the models using respiration and $SpO_2$, respiration-only, and $SpO_2$-only respectively. Correlation between expert-labelled and predicted AHI was 0.96, 0.78, and 0.93, respectively.

*Conclusions*: A wearable respiratory effort signal with or without SpO2 predicted AHI accurately. Given the large dataset and rigorous testing design, we expect our models are generalizable to evaluating respiration in a variety of environments, such as at home and in critical care.

[1] Department of Neurology, Massachusetts General Hospital, Boston, MA, USA
[2] Sleep & Health Zurich, University of Zurich, Zurich, Switzerland
[3] Present Address: Department of Neurology, Washington University School of Medicine in St. Louis, USA
[4] Division of Pulmonary and Critical Care Medicine, Massachusetts General Hospital, Boston, MA, USA
[5] Department of Anesthesiology, Critical Care, and Pain Medicine, Massachusetts General Hospital, Boston, USA
[6] MyAir Inc, Boston, MA, USA
[7] Department of Medicine, Division of Pulmonary, Critical Care & Sleep, Beth Israel Deaconess Medical Center, Boston, MA, USA

* Co-first authors, equal contribution,    † Co-senior authors, equal contribution.
Corresponding author: M. Brandon Westover, mwestover@mgh.harvard.edu

I. INTRODUCTION

Abnormalities of breathing during sleep may reflect both a primary sleep disorder (sleep apnea) or the health of the brain, lungs, or heart. There is broad utility if an easy to apply, minimally obtrusive and accurate measure of respiration is available, either for sleep apnea diagnostics or to track key vital signs in a range of situations. The latter includes intensive care units, heart failure and lung disease admissions or home tracking, and during conscious sedation. In such situations, abnormal breathing is expected, and could be used as a biomarker of recovery or treatment effects. Respiratory tracking may also be used to identify those at higher risk for post-operative delirium[1]. Central sleep apnea, which occurs when breathing stops during sleep because the brain fails to send signals to the muscles that control breathing[2] is especially common in hospitalized patients and may predict readmissions in systolic heart failure[3].

Obstructive sleep apnea is more common in the general population than central sleep apnea, and it is estimated that 2% to 4% of the adult population is affected[4]. A recent study on global prevalence of obstructive sleep apnea reports that almost one billion people are affected and stresses the urge to improve both healthcare and cost-effectiveness[5]. The standard method for sleep apnea detection is for patients to undergo a sleep study at a sleep centre or hospital. These sleep studies use polysomnography (PSG) to detect sleep apnea. PSG includes data from multiple signals involving many wires and is often uncomfortable for the patient. The recording from the night is then manually scored by trained sleep experts who detect apnea and hypopnea events and arousals. This process takes significant time and effort and is susceptible to errors. Even home sleep apnea tests are not readily repeatable, and current guidelines emphasize manual scoring of respiratory events.

Many studies recently have attempted to implement machine learning and deep learning techniques to automate the apnea-hypopnea detection process from PSG signals[6–9]. However, there is a need for a diagnostic tool for detecting sleep apnea that is suitable both for home settings as well as critical care.

The ideal respiratory monitoring device should have at least three characteristics: ease of use/application, automated analytics, and accuracy. In the present study we focus on a method for detecting sleep apnea, as a surrogate for the general capability of detecting abnormal respiration during sleep, from a practical wearable respiratory belt on its own or in combination with $SpO_2$. Our analytics approach consists of the extraction of clinically relevant and interpretable features, as well as a machine learning model (random forest). While the model predicts apnea events, the underlying respiratory features can give further insight to clinicians.

We show a novel, automated method for the detection of sleep apnea and AHI categorization with patient-friendly equipment suitable for a variety of clinical settings. We further provide the main code used in this study, ten PSGs with annotations and the wearable respiratory signals, and all trained models on our GitHub page[10].

II. MATERIAL AND METHODS

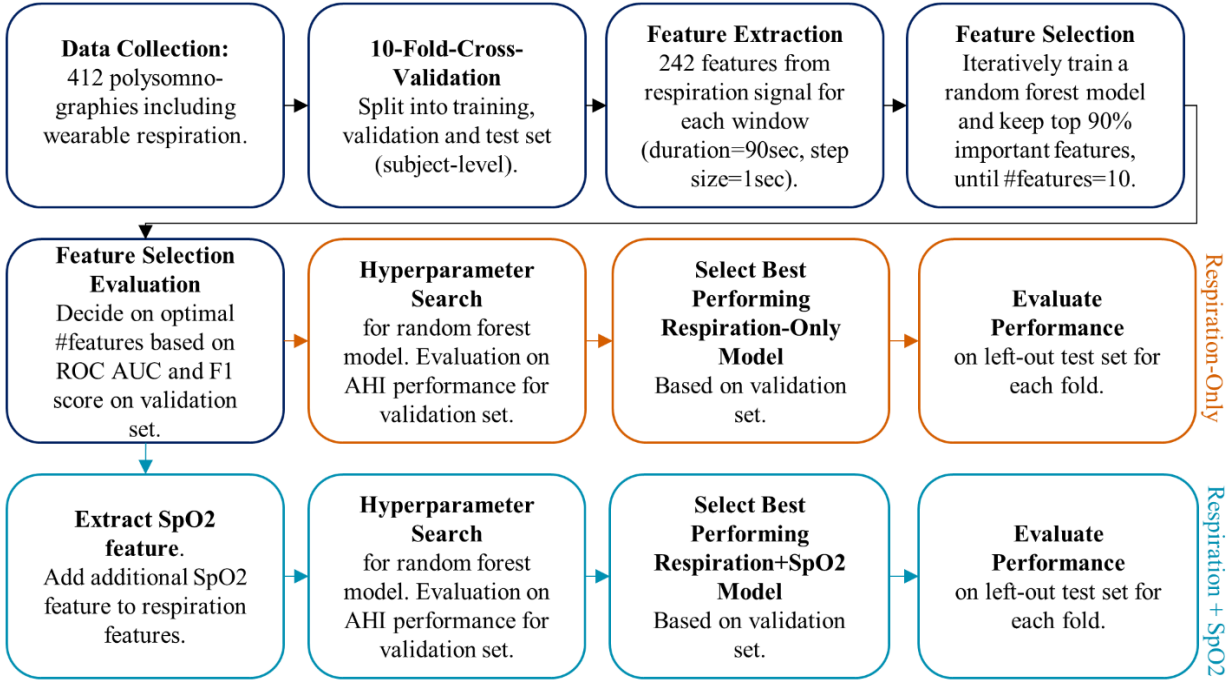

Figure 1. Flowchart of the modelling approach. We train models to detect apnoea events and predict Apnoea Hypopnea Index based on either the wearable respiration signal only, or on respiration and oxygen saturation signals.

*Data Collection*: The data, 412 PSGs (Natus System[11]) from 404 patients who also wore a wearable respiration belt (Airgo[12], see Figure E1 in appendix) were collected at the Massachusetts General Hospital sleep lab between January 2019 and January 2020 with approval of the Partners Institutional Review Board under protocol 2018P002937. Participants who were coming to the sleep lab for an overnight sleep study were enrolled through verbal consent shortly before the onset of PSG. Patients who agreed to participate wore the respiratory belt during their sleep study. There were no exclusion criteria. Experts at the sleep lab annotated the PSG recordings according to the American Academy of Sleep Medicine 2007 manual[13]. PSG and wearable recordings synchronization have been manually reviewed and corrected if necessary.

*Apnea and Hypopnea Scoring Rule:* There are two different accepted rules for scoring hypopneas according to the AASM manual[13], the '3% rule' and '4% rule', see appendix. The data was scored by experts (experienced registered polysomnographic technicians) using the 4% rule and we designed an automatic hypopnea scoring algorithm, using all PSG signals, to determine ground truth for the 3% rule and 4% rule.

We use the expert and automatic labels to train machine learning models on the wearable respiration and oximetry signals, resulting in the following ground truth labels: Expert Labels 4% Rule ('EL4'), Auto Labels 4% Rule ('AL4'), and Auto Labels 3% Rule ('AL3').

*Cross-Validation:* To obtain an unbiased performance estimate for the models, we split the data subject-wise into training, validation, and test sets in a 10-fold-cross-validation fashion, see appendix.

*Definition of Apnea Event:* All events labelled as any type of apnea or hypopnea are treated as an 'apnea event.' All remaining data are categorized as a 'non-apnea-event'.

*Overall Modelling Approach:* (see Figure 1 and Figure E2) We used a 90-second window, allowing enough context information before and after a typical 10 to 30 second lasting apnea event, and a step size of 1 second to extract features from the respiration signal, which were used to train a random forest model[14] for the binary apnea classification task ('Model Respiration-Only'). Furthermore, one feature from the $SpO_2$ signal was extracted (desaturation drop within 45 seconds of the detected event), and together with the respiration features was used to train another model with the same goal ('Model Respiration+$SpO_2$'). A third model was trained with a very similar approach to the Respiration+$SpO_2$ model, but the $SpO_2$ feature was extracted to be insensitive to a time lag of up to ±30 seconds between the respiration and $SpO_2$ signal ('Model Respiration+$SpO_2$ (robust)'). The latter is intended for scenarios where time alignment of signals coming from different sensors cannot be guaranteed. Finally, we constructed an '$SpO_2$-Only' model, that solely checks if the magnitude of a desaturation drop is above a certain threshold.

Table 1. Demographics

| Demographic | n (%) |
|---|---|
| Age (years)[a] | 56 (16) |
| Sex | |
|    Male | 220 (54%) |
|    Female | 182 (45%) |
|    Unknown | 2 (0.5%) |
| Race | |
|    White of Caucasian | 257 (64%) |
|    Black or African American | 20 (5%) |
|    Asian | 16 (4%) |
|    Native Hawaiian or Pacific Islander | 1 (0.2%) |
|    American Indian or Alaska Native | 1 (0.2%) |
|    Other or Unknown | 109 (27%) |
| Hispanic Ethnicity | 17 (14%) |
| Study Type | |
|    Diagnostic | 193 (47%) |
|    Split Night | 112 (28%) |
|    Full Night Titration | 106 (26%) |
| Apnea-Hypopnea-Index[a] | 11 (13) |
| Charlson Comorbidity Index[a] | 1.4 (1.6) |

[a] Mean (Standard Deviation)

The following four types of features were extracted at various timepoints (see Figure E3): standard deviation of the signal, sum of positive first derivative, sample entropy[15, 16], and Katz fractal dimension[17]. From a total of 242 features we performed feature selection[18, 19] and we determined the number of features that achieved an optimal median receiver operating characteristic curve (ROC AUC) [20] among the ten folds. Furthermore, we performed model hyperparameter optimization.

We used the models to predict apnea events and AHI, and evaluated the performances using ROC AUC, area under the precision recall curve (PRC AUC) [21], accuracy, sensitivity, precision, F1 score[22], and correlation between AHI and oxygen desaturation.

See appendix for further details on methodology.

## III. RESULTS

*Apnea and Hypopnea Scoring Rule:* The Pearson Correlation between the hypopnea indices (HI) based on expert and automatic PSG-based labels were 0.90 for Expert 4% and Auto 4%, 0.70 for Expert 4% and Auto 3%, and 0.84 for Auto 4% and Auto 3%, (see Figure E4). The mean HI are the same for the expert and auto 4% approaches, and there is a mean increase of 6 hypopneas per hour when using the 3% rule. These results (together with a sleep expert's manual review session of the 3%-rule auto-detections) show the applicability of the automatic hypopnea label algorithm (see Figure E5).

*Feature Selection and Feature Importance*: As illustrated in Figure E6, the individual folds resulted in AUCs between 0.8 and 0.92 and F1 scores between 0.55 and 0.75 on the validation sets. The mean and median performances increased up to approximately 40 features, plateaued until approximately 55, and then performances decreased again ('overfitting').

Analysing those results, we decided to use a maximum of 51 features (per fold) and exclude all sample entropy and fractal dimension features computed on the raw (non-smoothed) respiration signal, as the smoothed-version features showed higher importance.

The most important respiration features among all folds were (descending order):

1. Ventilation features at positions 55, 60, 65, 75, 80, 90 (all with width 10 seconds, and 12 second width for position 60).
2. Katz fractal dimension with width 10 seconds at positions 65 and 60.
3. Ventilation width 10 minutes ('reference')
4. Katz fractal dimension with width 60 seconds at position 90.
5. Standard deviation with widths 8 and 10 seconds at positions 60 and 80, and width 10-minute ('reference').
6. Sample entropy with width 30 seconds at positions 70 and 75.

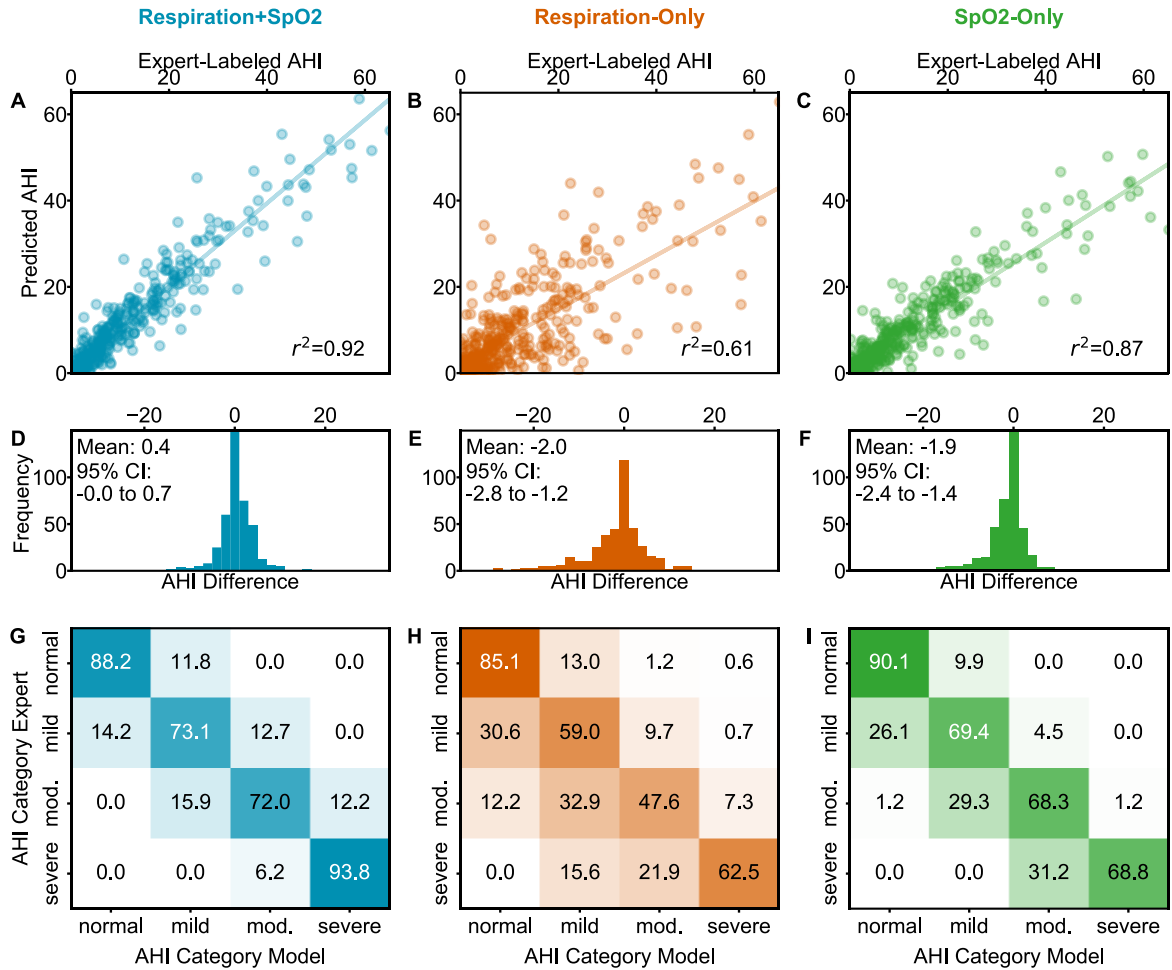

Figure 2. Apnoea Hypopnea Index (AHI)-based model performance evaluation (based on expert labels) for models trained on features from respiration and $SpO_2$ (left column), respiration only (middle column), and SpO2-only (right column). Results are obtained from left-out test data in a 10-fold cross-validation fashion; n=409 polysomnographic recordings including wearable respiration. Panels A-C: Scatterplots of predicted (model-based) and expert-labelled AHI per recording. Panels D-F: Difference of predicted and expert-labelled AHI, including 95% confidence interval (CI). Both model results show a unimodal error distribution. Panels G-I: Confusion matrix (in %) with AHI categorization (Normal: 0-5, Mild: 5-15, Moderate: 15-30, Severe: > 30), where rows: expert-labelled AHI category, columns: predicted AHI category. Accuracies for categorizations: 80% (Respiration+$SpO_2$), 67% (Respiration-Only), and 77% ($SpO_2$-Only).

*Event based:* Table E1 shows results for the event-based performance evaluation of the three models with different inputs. Adding the SpO$_2$ feature increased the overall sensitivity from 55.6% to 66.7% and reduced the number of false positives by nearly 50% to a false positive rate of 2.7%. The lag-robust version performed similarly to the 'non-robust' model and was just a fraction more sensitive (67.6% vs. 66.7%) and less precise. When including the SpO$_2$ feature, the largest increase in class-based sensitivity occurs for hypopneas, which is not surprising given that its definition includes 'a reduction of airflow of more than 50%' but not a complete absence of respiration – so this event was expected to be more difficult to detect looking at respiration signal only.

*Mean patient performance:* Model performances averaged over all PSGs that contained at least five expert-labelled apnoea events (n=360) (see Table 2 for Expert 4% and Auto 3% rules, and Table E2 for Auto 4% rule) showed:

1. The results are similar for all ground truth label versions (3% and 4%-hypopnea-scoring rule), showing the model approach is suitable for either of the scoring rules.
2. ROC AUC is greatest for Respiration+ SpO$_2$ models (0.94), followed by Respiration-Only (0.86) and SpO$_2$-Only (0.82).
3. PRC AUC is greatest for SpO$_2$-Only (0.51), followed by Respiration+ SpO$_2$ (0.48) and Respiration-Only (0.32).
4. The Respiration+SpO$_2$ robust version shows only a slight performance decrease compared to Respiration+ SpO$_2$ (ROC AUC 0.93 vs. 0.94, PRC AUC 0.45 vs. 0.48, respectively).

*Apnea Hypopnea Index:* The AHI evaluation from the Respiration+SpO$_2$ model resulted in coefficients of determinations ($r^2$) 0.92 and 0.85 for the Expert Labels (4%-hypopnea rule) and Auto labels (3%-hypopnea rule) respectively. The respective AHI categorization accuracies (Acc) were 0.8 and 0.7. See Table E3 in the appendix for $r^2$ and accuracy values for the Respiration-Only and SpO$_2$ only models.

Figure 2 shows the scatterplot for predicted and expert-labelled AHI for the Respiration+SpO$_2$, the Respiration-Only model, and SpO$_2$-Only Model (panels A,-C), the distribution of the absolute AHI difference (panels D-F) and the AHI-categorization confusion matrix (panel G-I). The coefficients of determinations of 0.92, 0.61 and 0.87 respectively show that all models can predict AHI with good agreement to the true AHI. We observed an increased performance if the SpO$_2$ feature is added together with the respiration signal. Figure E7 and Figure E8 show similar trends for the Auto 3% and Auto 4%-rule.

The ROC AUCs for the binarized AHI categorization tasks are between 0.94-0.98 (Respiration+ SpO$_2$), 0.85-0.95 (Respiration-Only), and 0.91-0.97 (SpO$_2$-Only) for the different hypopnea rules, indicating that the models accurately distinguish between AHI categories for any input signal and hypopnea rule used (see Figure 3).

Example model predictions and expert labels are shown together with the respiration and oxygen saturation signals in Figure 4.

*SpO$_2$ burden:* The correlations between the ground truth labels and SpO$_2$ labels are between 0.46-0.53. For the Respiration+SpO$_2$ models, the correlations are between 0.50-0.52, and for the Respiration-Only model they are between 0.37-0.39. See Table E4.

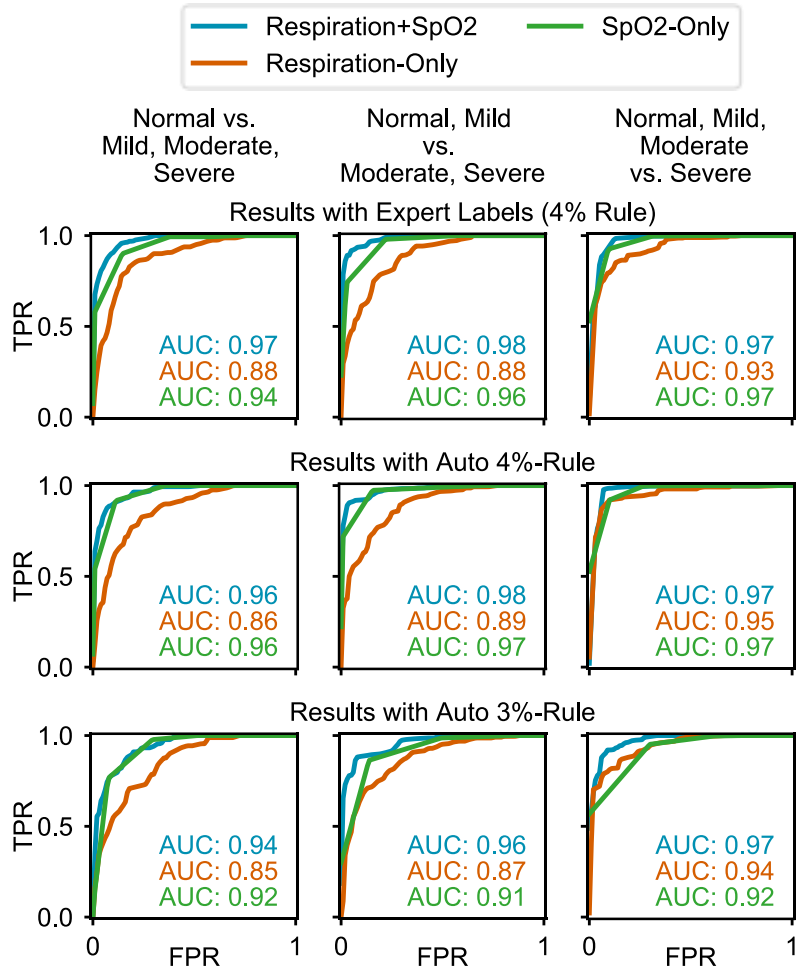

Figure 3. Receiver Operating Curves and the corresponding areas under the curve (AUC), evaluated on binary AHI categorization tasks, for different hypopnea scoring rules. All models result in AUCs greater than 0.85, showing an accurate AHI categorization.

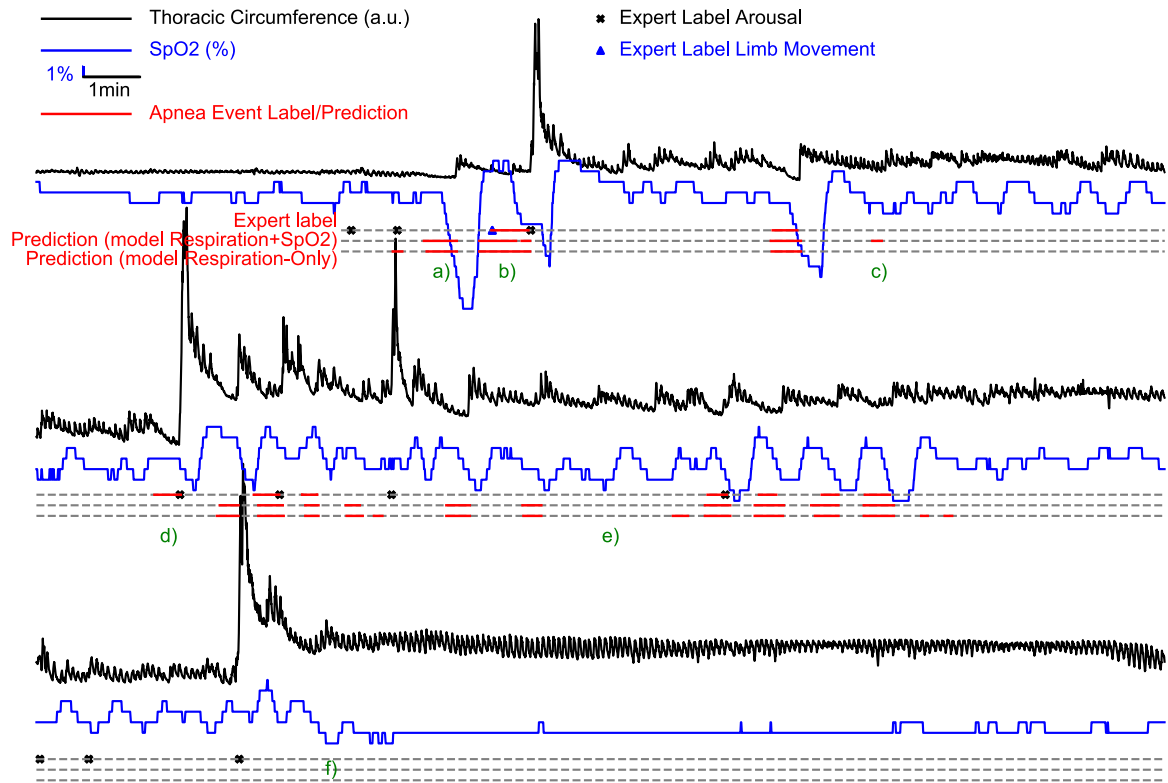

Figure 4. Example signals and predictions: 60 minutes of a recording from a 47-year old subject, where the AHI prediction is comparable to the mean model performance (true AHI: 7, predicted AHI respiration+SpO$_2$:8, predicted AHI respiration-only: 10). The respiration and SpO$_2$ signals as well as the apnoea events annotated by the sleep experts and predicted by our two models are presented. Point a) shows an instance where both models labelled an apnoea event, but the sleep experts did not, even though a decrease in the respiration signal and a large drop in blood oxygenation is visible. Such instances are treated as false positives, even though this event is either missed by the sleep experts or the event potentially narrowly fails to qualify for an apnoea event. Point b) is an example of an event where the sleep experts and both models label an apnoea event. Point c) shows an instance where the Respiration+SpO$_2$ model incorrectly labels an apnoea event, but the Respiration-Only model does not. Point d) is labelled as an apnoea event by the sleep expert but is not detected by either model (false negative). The region around point e) shows a series of similar events, all showing a cyclic decrease and increase in respiration amplitude and desaturation. Some of those events are marked as apnoea events by the sleep expert, some are not – illustrating the difficulty for a model to fully agree with the human labels. At point f), the respiration clearly switches from an instable period to a more stable period and it becomes more regular – neither the sleep expert nor the models detect events here.

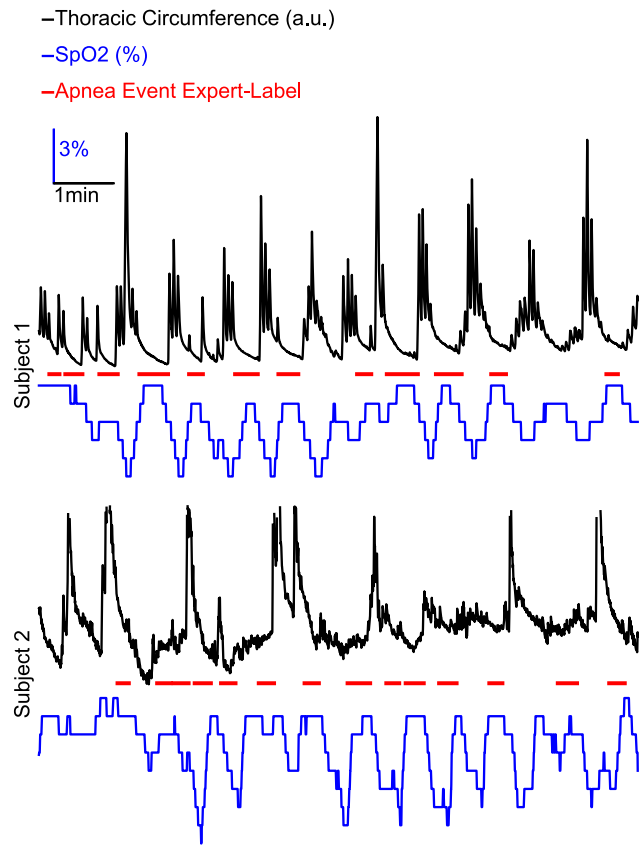

Figure 5. 10-minute respiration and SpO2 signal tracings from two subjects that have a very similar expert-scored AHI (15 and 17) but clearly show different kinds of apnoea events. These different phenotypes are reflected in the obstructive apnoea indices (OI) and central apnoea indices (CI), subject 1: OI=1, CI=7, subject 2: OI=12, CI=2. The wearable respiration signal not only contains information to automatically assess the AHI, but to differentiate phenotypes.

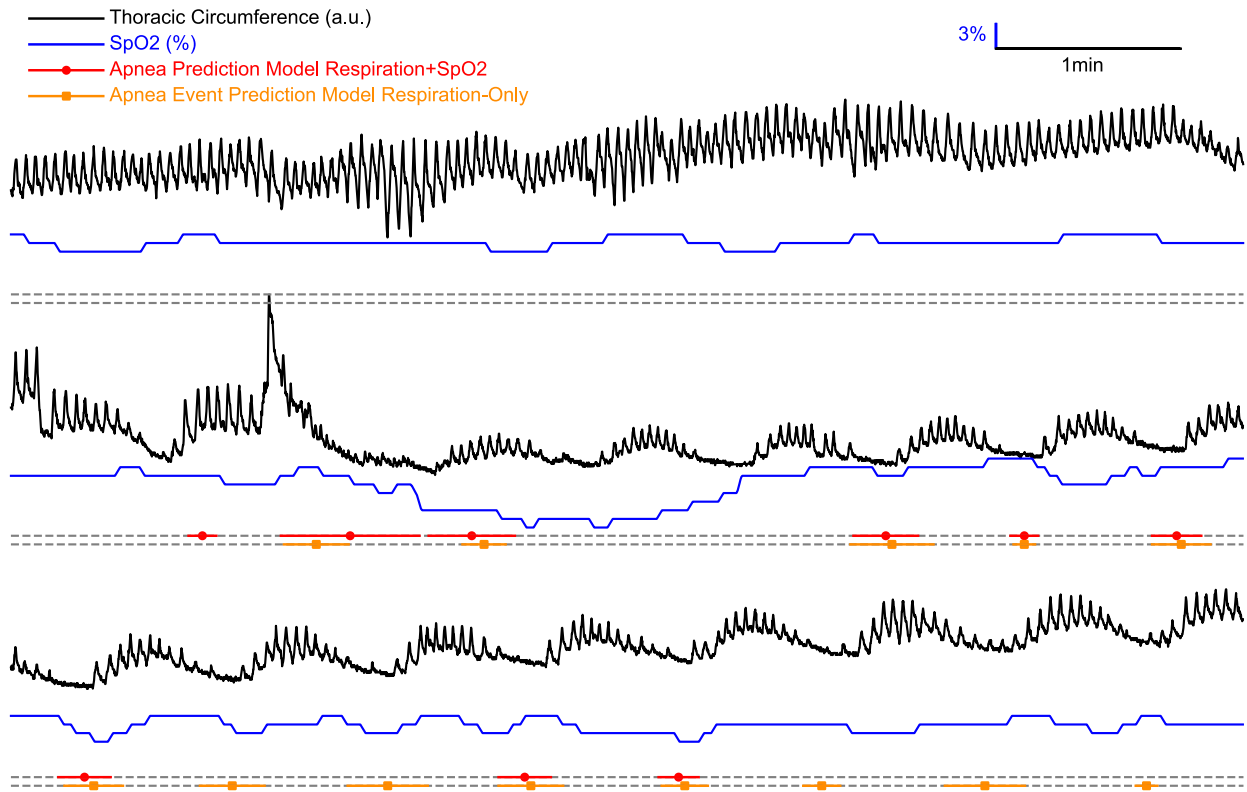

Figure 6. Example signals and predictions: 20 minutes of a recording from an 88-year old woman who was admitted to the ICU for a hip fracture. The patient has a Charlson Comorbidity Index of 0, and no previous obstructive sleep apnea diagnosis. The apnea predictions from the Respiration+SpO2 and Respiration-Only models are shown on the figure. The detected AHI from the Respiration+SpO2 model was 9.6, and from the Respiration-Only model it was 12.7. The average SpO2 value was 98.0 ± 1.6, with a maximum value of 100.0 and a minimum of 92.0.

Table 2. Mean Patient Performance

| Performance Metric | Respiration+SpO$_2$ | Respiration+SpO$_2$(robust) | Respiration-Only | SpO$_2$-Only |
|---|---|---|---|---|
| Expert Labels (4% Hypopnea Rule) | | | | |
| ROC AUC | 0.94 (0.93 to 0.94) | 0.93 (0.92 to 0.93) | 0.86 (0.85 to 0.87) | 0.82 (0.81 to 0.83) |
| PRC AUC | 0.48 (0.46 to 0.50) | 0.45 (0.42 to 0.47) | 0.32 (0.30 to 0.35) | 0.51 (0.50 to 0.53) |
| Accuracy | 0.94 (0.93 to 0.95) | 0.94 (0.93 to 0.95) | 0.92 (0.92 to 0.93) | 0.96 (0.95 to 0.97) |
| Sensitivity | 0.58 (0.56 to 0.61) | 0.58 (0.56 to 0.61) | 0.50 (0.48 to 0.53) | 0.73 (0.71 to 0.76) |
| Precision | 0.50 (0.48 to 0.53) | 0.44 (0.42 to 0.46) | 0.32 (0.30 to 0.34) | 0.57 (0.55 to 0.58) |
| F1 Score | 0.52 (0.50 to 0.54) | 0.48 (0.46 to 0.50) | 0.36 (0.34 to 0.38) | 0.62 (0.61 to 0.64) |
| Automatic Labels (3% Hypopnea Rule) | | | | |
| ROC AUC | 0.93 (0.92 to 0.93) | 0.92 (0.92 to 0.93) | 0.86 (0.85 to 0.87) | 0.83 (0.82 to 0.83) |
| PRC AUC | 0.44 (0.42 to 0.46) | 0.42 (0.40 to 0.44) | 0.32 (0.30 to 0.34) | 0.52 (0.51 to 0.53) |
| Accuracy | 0.95 (0.94 to 0.96) | 0.95 (0.94 to 0.96) | 0.93 (0.93 to 0.94) | 0.95 (0.94 to 0.96) |
| Sensitivity | 0.51 (0.48 to 0.53) | 0.54 (0.51 to 0.56) | 0.53 (0.51 to 0.56) | 0.85 (0.83 to 0.87) |
| Precision | 0.52 (0.50 to 0.54) | 0.48 (0.45 to 0.50) | 0.36 (0.34 to 0.38) | 0.49 (0.46 to 0.51) |
| F1 Score | 0.50 (0.48 to 0.52) | 0.49 (0.47 to 0.51) | 0.40 (0.37 to 0.42) | 0.59 (0.57 to 0.61) |

## IV. DISCUSSION

Detecting abnormal sleep respiration including classic sleep apnea in a patient-friendly and low-cost manner can enable clinicians and researchers to better understand and treat primary sleep disorders and abnormal sleep-breathing in a range of medical and neurological conditions. The present study provides new measurement insights in this area as it:

a) uses a large dataset of simultaneously recorded wearable signals and polysomnographies (as the gold standard) with more than 400 subjects.
b) uses a rigorous method of model performance evaluation, i.e. no information from the test patients was used for model training (cross-validation), making the reported performance an unbiased estimate of what one can expect when using this model in the future on a similar population.
c) shows state-of-the-art performance in automatically assessing AHI, compared to other studies published using wearable or portable signals (see Table E5). Performance comparison with other studies is challenging, as even when the same performance metrics (such as accuracy) are used, the underlying approach to compute the metrics might differ. Moreover, a substantial number of studies have not reported a complete view of the performance; for example, reporting sensitivity- and specificity-based metrics but not precision-based metrics, or reporting event-based metrics but not patient-based (AHI) metrics.
d) investigates the question of how much it helps in predicting AHI to have an SpO$_2$(%) sensor in addition to the wearable respiratory device, where we observed an increase in performance for most metrics. This leads us to the conclusion that having both sensors will facilitate the most reliable automatic AHI assessment and further analysis, such as apnea phenotyping.

Moreover, for real-world (and non-ideal sleep lab settings), we showed that allowing a time lag between the respiration and SpO$_2$ signals (potentially recorded from two different sensors) of up to ±30 seconds, performance is practically unchanged.

In this paper we have shown that it is possible to detect apnea events using a single, wearable respiratory belt with or without SpO$_2$, and from SpO$_2$ alone. While we show here that it is possible to accurately compute AHI from SpO$_2$ signals alone, it is important to note that the respiration signals from this wearable respiratory device provide additional important phenotypic information about breathing patterns that one may not obtain from conventional SpO$_2$ metrics alone. Figure 5 shows an example of two distinctly different phenotypic breathing patterns from two patients with similar AHI score. This illustrates the importance of measuring respiration when detecting sleep apnea or tracking sleep-breathing in diverse conditions, in addition to SpO$_2$ information. We have also shown that this model for apnea and hypopnea detection is robust. We tested the model using both of the hypopnea scoring rules recommended by the AASM manual[13], and the results for each rule showed accurate AHI prediction.

Our results have implications for both sleep apnea diagnostics, and sleep-breathing tracking in environments such as the intensive care unit. There are several approaches to monitoring respiration in the intensive care unit [23, 24], including pulse oximetry, ventilator pressure and flow waveforms, lung volume measurement, electrical impedance tomography, respiratory information embedded in hemodynamic monitoring, volumetric capnography, esophageal and transpulmonary pressures and diaphragmatic electromyography. Each technique has advantages and disadvantages, and some cannot be continued once mechanical ventilation is stopped. The technology we describe is simple, and can move with the individual patient, including from hospital to home. Sleep based on electroencephalography is notoriously difficult to measure in the intensive care environment, and oximetry typically is used as a safety measure. Technologies and analytics of the type we describe could perform as an enhanced vital sign, including post-weaning from the ventilator, when patients are especially vulnerable. The effect of treatment of heart failure or lung disease exacerbations could also be tracked.

In the context of sleep apnea, a diagnosis from a wearable respiration device with or without the addition of SpO$_2$ allows greater flexibility of sleep apnea diagnosis in cases where full PSGs or event home testing are not readily feasible. High resolution respiratory effort monitoring also enables extraction of sleep stage information[25], and sleep-breathing phenotype data, which may further enable improved precision management.

The concept of home sleep and sleep apnea testing continues to evolve rapidly with advances in technology. Current guidelines to assess obstructive sleep apnea do not include single respiratory effort belts [26]. Our results show that we can obtain clinically useful accuracy with a sensitive respiratory effort signal alone, combined with advanced analytics.

Some limitations to the study include potential selection bias from the fact that the study population was chosen from a single-centre and included only patients already coming to the sleep lab for testing. Thus, we did not have hospitalized and critically ill patients, where breathing abnormalities may be harder to quantify. Including more patients from multiple hospitals and patients from the general public could have increased the breadth of the study population and the generalizability of the findings. Additionally, only one sleep expert scored each PSG recording, as is typical of clinical recordings. Although there are seven sleep experts (technicians) at the centre who score PSGs, each individual recording is only scored by one of the experts. Therefore, there is a possibility that that there is some variation or label noise when it comes to the ground truth sleep expert labels, because scoring PSGs is a subjective process and can be biased based on the expert that scored the recording. This concern is largely mitigated by the finding that results were similar when evaluating the model against ground truth labels determined from the PSG by automated methods.

## V. Conclusions

Respiratory effort signals obtained with a single patient-friendly wearable band, standard oxygen saturation signals, and an automated analytics approach allow us to extract clinically relevant respiratory features and detect sleep apnea with high accuracy.

In future work, we plan to apply this model to ICU patients. Patients enrolled in a clinical trial [27] in the ICU at Massachusetts General Hospital wear a respiratory belt for the study, and we plan to use this information in addition to $SpO_2$ to detect apneas in this population of patients. For the patients already enrolled, we performed a preliminary check using our models and obtained promising initial results – Figure 6 shows the resulting apnea predictions for one example patient.

Compliance with Ethical Standards:

Funding: This study was funded by the NIH (1R01NS102190, 1R01NS102574, 1R01NS107291, 1RF1AG064312).

Conflict of Interest:

Wolfgang Ganglberger declares that he/she has no conflict of interest.
Abigail A. Bucklin declares that he/she has no conflict of interest.
Ryan A. Tesh declares that he/she has no conflict of interest.
Madalena Da Silva Cardoso declares that he/she has no conflict of interest.
Haoqi Sun declares that he/she has no conflict of interest.
Michael J. Leone declares that he/she has no conflict of interest.
Luis Paixao declares that he/she has no conflict of interest.
Ezhil Panneerselvam declares that he/she has no conflict of interest.
Elissa M. Ye declares that he/she has no conflict of interest.
B. Taylor Thompson reports personal fees from Bayer and Thetis, outside the submitted work.
Oluwaseun Akeju declares that he/she has no conflict of interest.
David Kuller reports non-financial support from Myair Inc., during the conduct of the study; non-financial support from Myair Inc, outside the submitted work; In addition, Dr. Kuller has a patent Patent US10123724B2 "Breath volume monitoring system and method" issued.
Dr. Thomas reports personal fees from GLG Councils, Guidepoint Global, and Jazz

Pharmaceutics, outside the submitted work. In addition, Dr. Thomas has a patent ECG-spectrogram with royalties paid by MyCardio, LLC, a patent Auto-CPAP with royalties paid by DeVilbiss-Drive, and an unlicensed patent $CO_2$ device for central / complex sleep apnea issued. Dr. Westover reports grants from NIH, during the conduct of the study.

Ethical approval: All procedures performed in studies involving human participants were in accordance with the ethical standards of the institutional and/or national research committee and with the 1964 Helsinki declaration and its later amendments or comparable ethical standards.

Informed consent: Informed consent was obtained from all individual participants included in the study.

# Appendix

METHODS

*Apnea and Hypopnea Scoring Rule:* According to the AASM manual [i], there are two different accepted rules for scoring hypopneas. Both rules have the same requirement for airflow decrease (measured on nasal pressure or PAP device flow signals) but differ in that a) there must be at least 3% oxygen desaturation from baseline or the airflow decrease is associated with an arousal, or b) there is at least 4% oxygen desaturation. For simplicity, we call the first rule '3% rule' and the second '4% rule'. The sleep lab dataset used for this study was scored by "experts" (registered and experienced polysomnographic technicians) using the 4% rule. At the time of data collection, the 3% rule was not being implemented. To create apnea prediction models that were suitable with the 3% rule as well, we designed the following automatic hypopnea scoring algorithm to determine ground truth for the 3% rule based on full PSG data:

1) Compute airflow decrease and check if requirement, as specified in the AASM manual, is fulfilled.
2) Compute the oxygen desaturation and check if a 3% threshold is exceeded.
3) Check if the expert has labelled an event as an arousal.
4) If 1) and either 2) or 3) is fulfilled, an event is classified according to the 3% rule.

We then use these automatic labels derived from the PSG data in the same manner as the expert labels, i.e. as ground truths to train machine learning models that act on the reduced data. We computed the auto-labels for both the 3% and the 4%-rule, the latter to validate the automatic approach, as the auto-hypopnea-labels with the 4% rule should be similar to the expert hypopnea labels. We computed pairwise Pearson correlation coefficients between the hypopnea indices (#hypopnea/#hours of sleep) of all patients for all three labeling methods.

*Cross Validation:* In each fold, recordings from 80% of all unique subjects were used for training, while both the validation and test sets consisted of 10% of patients each. For patients with multiple recordings, all recordings from the same patient were in the same set.

*Signal Preprocessing:* The only preprocessing step done in the respiration signal was normalization: For each recording, the mean µ and standard deviation σ of the 0.01 and 0.99-quantiles clipped signal was used to normalize the signal s: s = (s- µ)/ σ.

*Feature Extraction*: Twelve timepoints of interest were defined for each 90-second segment:10, 20, 30, 40, 45, 50, 55, 60, 65, 70, 80, and 90 seconds (see E3 for an illustration). Four types of features were extracted at these timepoints:

1. Standard deviation of the signal for the past 8, 10, and 12 seconds; we also added past 5-min and 10-min (long-term) references.
2. Sum of positive first derivative (as an estimate of ventilation) for the past 8, 10, and 12 seconds and additionally a 5-min and a 10-min feature.
3. Sample Entropy[ii]: for the past 10, 30, 45, and 60 seconds.
4. Katz Fractal Dimension[iii]: for the past 10, 30, 45, and 60 seconds.

Features 1 and 2 were aimed at encoding the information about the expected decrease in standard deviation and ventilation during an apnea period compared to before and after an event. Features 3 and 4 were aimed at encoding information about the regularity of the respiration, as we expected periods with apneas have a more irregular, unpredictable pattern than normal breathing. Features 3 and 4 were computed both from the raw respiration signal and from a 1.2-second-moving average of the respiration signal that is supposed to be less affected by short, small artifacts such as movement and heartbeats. In total, those four feature categories with different look-back times and positions in the 90-second window led to 242 features.

*Feature Selection:* To reduce the number of features and avoid problems due to the curse of dimensionality[iv] and overfitting[v] the following iterative feature selection approach was designed:

1. All 242 features were computed for the full data.
2. Hierarchical clustering with Ward's method[vi] was used to group the features into clusters so that colinear features are in the same cluster.
3. The remaining features (first iteration: all features) were used to train a random forest model to predict apnea-events.
4. Feature permutation importance[vii] was computed, and the bottom 10% of features were removed. If all features from a cluster were in the bottom 10%, then one of the features from the cluster was kept. Therefore, all features from a cluster could only be removed if at a certain step one (and only one) feature was left in the cluster and was selected in the bottom 10%. This cluster-elimination behavior was used to account for colinear features that might have been ranked with low importance by permutation (because the colinear feature contains similar information) but in fact had high importance as a group.
5. Steps 3 and 4 were performed until there were ten features left.

To reduce computational complexity of feature computation, the sample entropy and Katz fractal dimension features were only kept for either the raw or 1.2-second-moving average version after result analysis, i.e. the versions that had the least surviving features in the remaining 10 features (in all folds) were eliminated.

To decide on the number of features to use, we evaluated the apnea-event prediction performance of all models, for all steps in the feature selection procedure, on the validation set. Performance metrics used were the area under the receiver operating characteristic curve (ROC AUC)[viii] and the F1 score (harmonic mean of sensitivity and precision)[ix]. We selected the number of features that had the best median AUC among all ten folds.

*Model output smoothing rule:* The model was designed to be applied to the data with a step size of one second, and to yield an event prediction each second. We applied a post-processing prediction output smoothing rule that grouped any potentially disjoined 1-second-predictions into joined event predictions: If, in a segment of 10 seconds, there were more than or equal to *i_positive_predictions* (hyperparameter, see below) apnea predictions on a 1-second resolution, then the whole 10-second segment was labelled as an apnea prediction. If this rule was not fulfilled, the whole 10-second segment was labelled as no-apnea prediction.

*Model Building and Hyperparameter Search:* With the number of features fixed, we then aimed to find a good set of hyperparameters for the binary classification random forest models (hyperparameter tuning used training data only). This step was done independently for the Respiration-Only and the Respiration+$SpO_2$ models. Hyperparameters analyzed were:

1. minimum numbers of samples needed for a split
2. imbalance of apnea-event and non-apnea-event samples
3. class-weights for apnea-events and non-apnea-events.

For a resulting set of models that performed best on the ROC AUC on the validation set, two further post-processing parameters were refined:

1. prediction threshold of classifier
2. prediction smoothing rule (*i_positive_predictions*)

A final set of hyperparameters was selected that offered the most favorable performance in terms of mean AHI prediction on the validation set.

*Model evaluation:* The models' performances were evaluated in the following ways:

1. Event-based: All test set recordings were pooled together and the number of true positives, false positives, false negatives, and true negatives were computed [x], where:

    -True positive: An expert-labelled event that is predicted by the model too, so that at least 50% of the expert-label overlaps with the prediction label.

    -False negative: An expert labelled event that is not predicted as an event by the model, i.e. no prediction or a prediction that overlaps less than 50% of the expert-label.

    -False positive: A model predicted event that is not expert-labelled, or less than 50% of the prediction's duration is expert-labelled as an event.

    -Number of true negatives = (seconds of data without any labelled or predicted event)/18. The median length of a labelled event in our dataset is 18 seconds. By normalizing the non-labelled part by the median, a proportionate number of true negative events is computed.

Furthermore, we computed the sensitivities of the different expert-labelled apnea events, e.g. sensitivity obstructive apnea = percentage of all expert-labelled obstructive apneas that were predicted as an event by the model.

2. Mean patient performance: With the event-based definitions above, we computed the following standard performance metrics for each recording independently and report the sample mean and associated 95% confidence interval: ROC AUC, area under the precision recall curve (PRC AUC)[xi], accuracy, sensitivity, precision, and F1 score. To avoid biasing the mean with recordings with nearly no events labelled (performance metrics can be very low or very high by chance in this case), recordings with less than 5 total expert-labelled events were excluded (this exclusion criteria affected 49 recordings and did not change the result significantly).
3. Apnea Hypopnea Index: Following AASM 2007 manual[xii], AHI was computed both for the ground truth labels, and for the model predicted apnea events: AHI = #of events / (hours of sleep).
   For computing the predicted AHI, predicted events during the "awake" sleep stage were excluded. The coefficient of determination ($r^2$) was computed as the squared Pearson correlation coefficient. For each recording, the difference of AHI prediction was computed, i.e. AHI_ difference = AHI_predicted – AHI_true, and visualized as a histogram with bin size 2.
   Next, we categorized each expert-labelled and predicted AHI according to the AASM manual's suggestion:
   
   | | |
   |---|---|
   | Normal: | 0-5 |
   | Mild: | 5-15 |
   | Moderate: | 15-30 |
   | Severe: | 30 or more. |

   A normalized confusion matrix of the categorized AHI prediction was computed and visualized (e.g. row 'mild', column 'severe' shows the percentage of all recordings that were categorized as 'mild' by the expert but 'severe' by the model), and the corresponding accuracy was derived. Furthermore, to evaluate the models' ability to discriminate between binary AHI categorizations, we compute ROCs and AUCs for the following classification tasks:

   a) normal vs. mild, moderate and severe.
   b) normal and mild vs. moderate and severe.
   c) normal, mild and moderate vs. severe.

To compute the ROCs, we vary the prediction threshold of the model. For each threshold, the predicted (binarized) AHI category for each patient is computed, from which a pair of sensitivity and specificity values is obtained.

SpO$_2$ burden: To quantify the blood oxygenation burden of a full night's sleep, we defined: SpO$_2$ burden = area under the median SpO$_2$ line / time, where only data during sleep was considered. The Pearson correlation between the AHIs (true and predicted) and the SpO$_2$ burden was computed.

Data analysis was done using Python 3.6[xiii], with main packages used: NumPy[xiv], pandas[xv], SciPy[xvi], Matplotlib[xvii], scikit-learn[xviii] and EntroPy[xix].

FIGURES AND TABLES

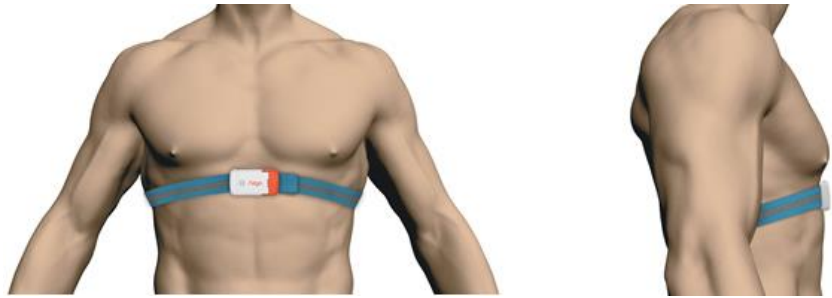

Figure E1. The wearable respiration device ('Airgo' [11]) used in this study

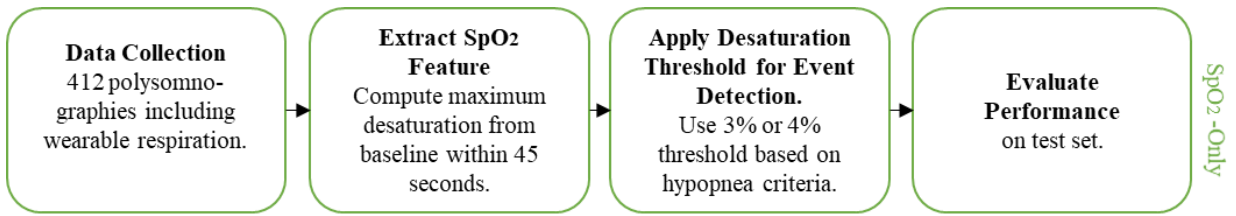

Figure E2. Flowchart of the modelling approach for the SpO2-Only model.

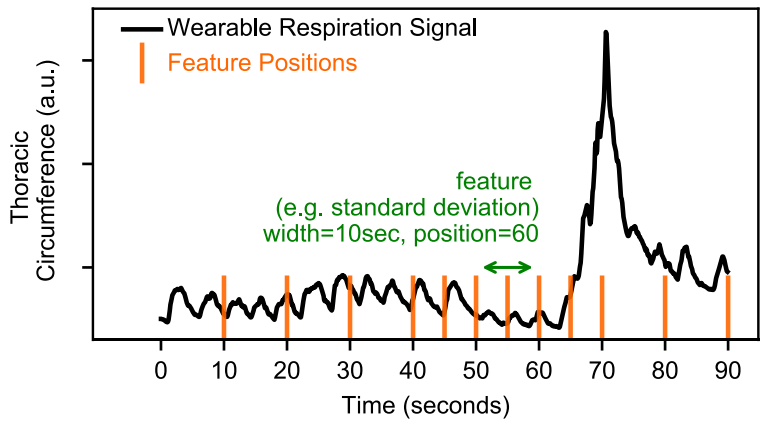

Figure E3. Illustration of feature extraction. For each window of 90 seconds (and step size one second) features like standard deviation, ventilation, fractal dimension, and sample entropy are extracted at twelve positions. The signal input to the feature ('width') varies from 10 seconds to 60 seconds (and 5 and 10 minutes for reference).

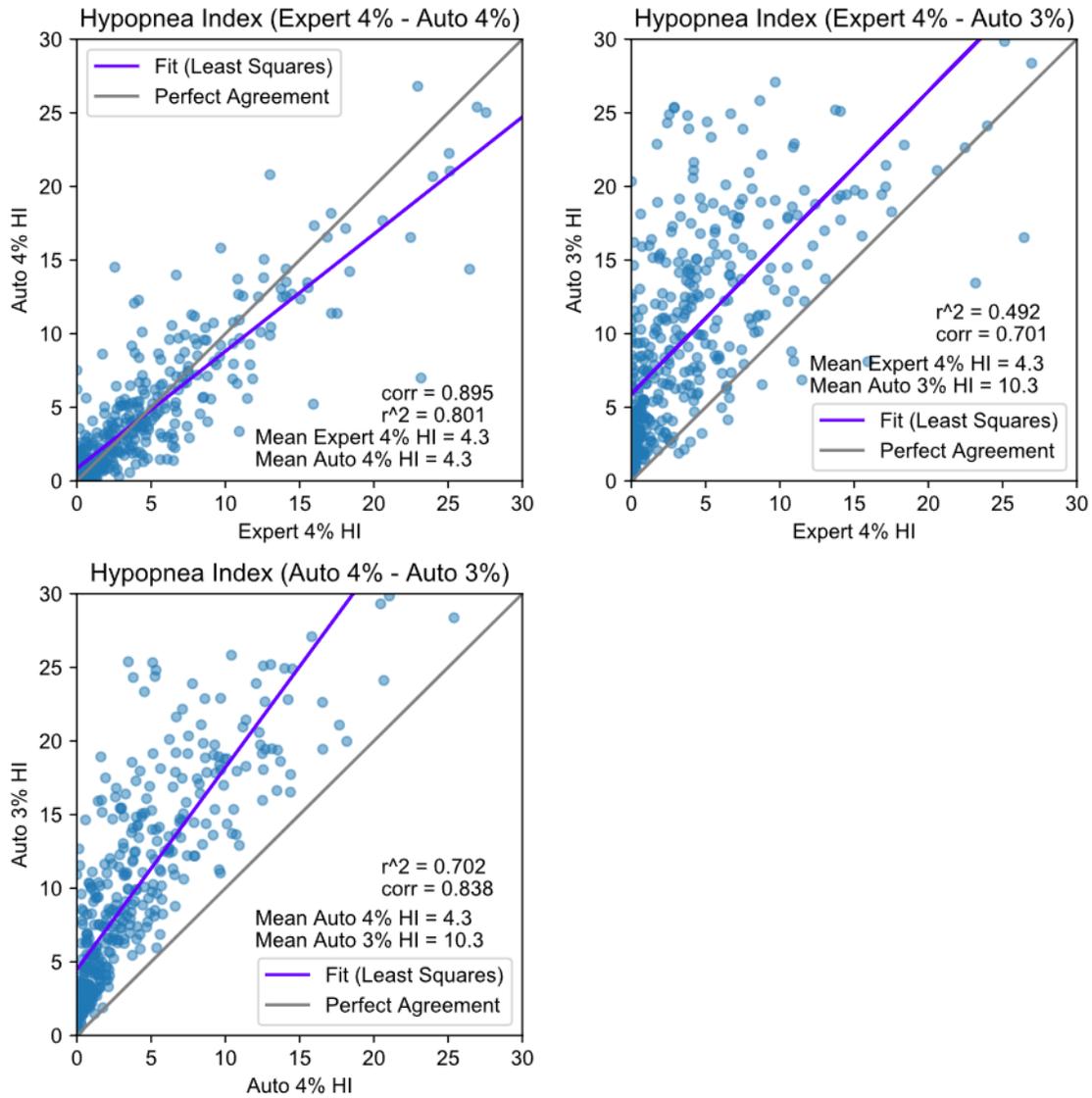

Figure E4. Automatic hypopnea algorithm results. The hypopnea labels of the expert and the auto 4% approach show a large agreement with a Pearson correlation of 0.9 and same mean hypopnea index (HI). Using the 3% rule leads to a mean increase of 6 hypopneas per hour

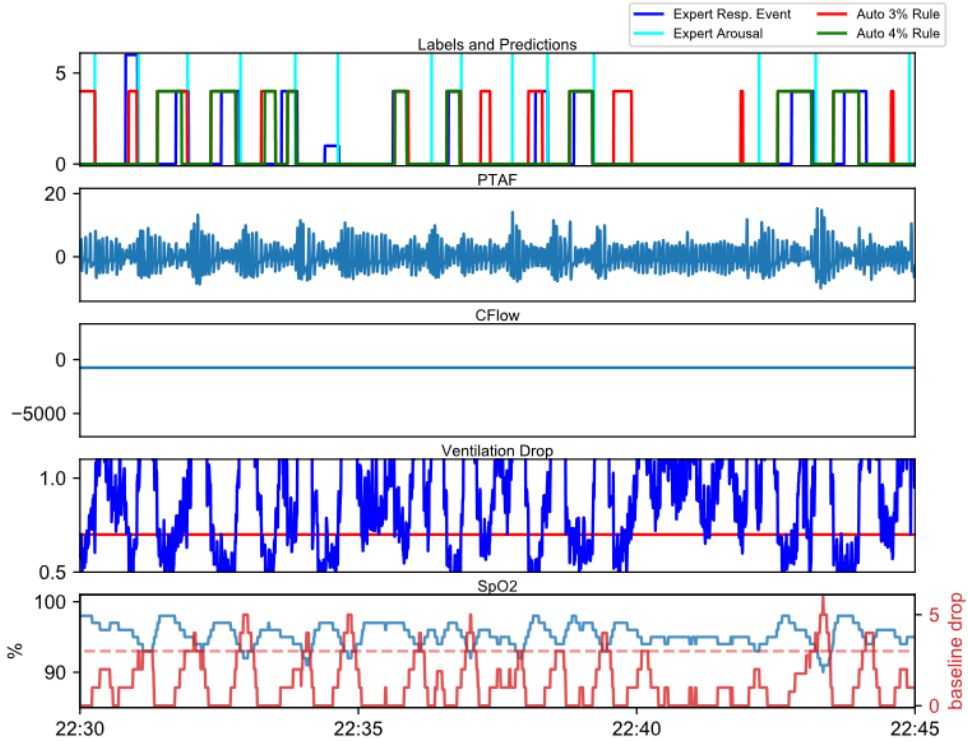

Figure E5. Automatic Hypopnea detection (and rule conversion) algorithm. Example results for a 15-minute signal during diagnostic part of a split night study (i.e. PTAF signal available). Colored rectangles with height 4 represent hypopnea labels for automatic and expert approach, blue rectangle with height 1 represents an event labelled as obstructive apnea by the expert. The fourth and fifth panel show the computed features from the airflow and SpO2 signal respectively that are used to assess if a threshold (as specified by American Society of Sleep Manual) is met.

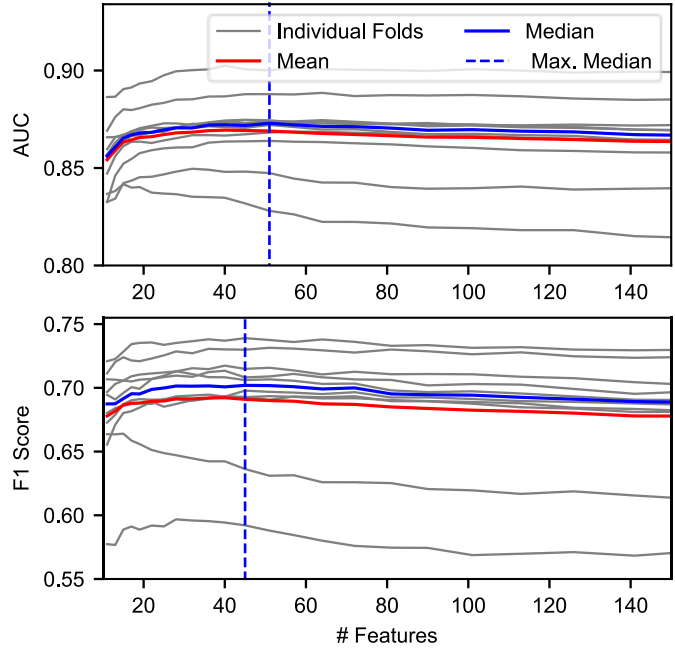

Figure E6. Results of iterative feature selection approach on the validation set for the 10 folds. A maximum of 50 features was selected as optimal.

TABLE E1. Performance Results Event-Based
PERFORMANCE EVALUATION OF THE MODELS INCLUDING THE WEARABLE RESPIRATION ON THE FULL DATA OF 409 POLYSOMNOGRAPHIC RECORDINGS (10-FOLD-CROSS-VALIDATION), USING THE EXPERT LABELS AS GROUND TRUTH.

|  | RESPIRATION+$SpO_2$ | RESPIRATION+$SpO_2$ (ROBUST) | RESPIRATION-ONLY |
|---|---|---|---|
| TOTAL HOURS SLEEP | 2,522 | 2,522 | 2,522 |
| ANNOTATED EVENTS | 27,592 | 27,592 | 27,592 |
| TRUE POSITIVES | 18,411 | 18,647 | 15,334 |
| FALSE POSITIVES | 14,458 | 18,460 | 28,555 |
| FALSE NEGATIVES | 9,181 | 8,945 | 12,258 |
| TRUE NEGATIVES | 517,269 | 518,738 | 524,442 |
| SENSITIVITY OBSTRUCTIVE APNEA | 0.68 | 0.7 | 0.55 |
| SENSITIVITY CENTRAL APNEA | 0.67 | 0.71 | 0.75 |
| SENSITIVITY MIXED APNEA | 0.54 | 0.56 | 0.57 |
| SENSITIVITY HYPOPNEA | 0.66 | 0.64 | 0.38 |
| PRECISION APNEA EVENT | 0.56 | 0.5 | 0.35 |

Table E2. Mean Patient Performance

| Performance Metric | Automatic Labels (4% Hypopnea Rule) | | | |
|---|---|---|---|---|
| | Respiration+$SpO_2$ | Respiration+$SpO_2$ (robust) | Respiration-Only | $SpO_2$-Only |
| ROC AUC | 0.93 (0.93 to 0.94) | 0.93 (0.92 to 0.93) | 0.86 (0.85 to 0.87) | 0.82 (0.81 to 0.83) |
| PRC AUC | 0.47 (0.45 to 0.49) | 0.44 (0.41 to 0.46) | 0.32 (0.30 to 0.34) | 0.51 (0.50 to 0.53) |
| Accuracy | 0.95 (0.94 to 0.96) | 0.95 (0.94 to 0.96) | 0.94 (0.94 to 0.94) | 0.96 (0.95 to 0.97) |
| Sensitivity | 0.51 (0.49 to 0.54) | 0.52 (0.50 to 0.55) | 0.50 (0.47 to 0.53) | 0.73 (0.71 to 0.76) |
| Precision | 0.59 (0.57 to 0.61) | 0.53 (0.51 to 0.55) | 0.40 (0.38 to 0.42) | 0.57 (0.55 to 0.58) |

TABLE E3 – AHI EVALUATION PERFORMANCE

| | EXPERT LABELS (4%-HYPOPNEA RULE) | | |
|---|---|---|---|
| | RESPIRATION+$SpO_2$ | RESPIRATION-ONLY | $SpO_2$ |
| R | 0.96 | 0.78 | 0.93 |
| $R^2$ | 0.92 | 0.61 | 0.87 |
| ACCURACY | 0.80 | 0.67 | 0.77 |
| | AUTO LABELS (4%-HYPOPNEA RULE) | | |
| | RESPIRATION+$SpO_2$ | RESPIRATION-ONLY | $SpO_2$ |
| R | 0.95 | 0.79 | 0.92 |
| $R^2$ | 0.91 | 0.62 | 0.85 |
| ACCURACY | 0.81 | 0.64 | 0.82 |
| | AUTO LABELS (3%-HYPOPNEA RULE) | | |
| | RESPIRATION+$SpO_2$ | RESPIRATION-ONLY | $SpO_2$ |
| R | 0.92 | 0.77 | 0.90 |
| $R^2$ | 0.85 | 0.59 | 0.81 |
| ACCURACY | 0.70 | 0.54 | 0.68 |

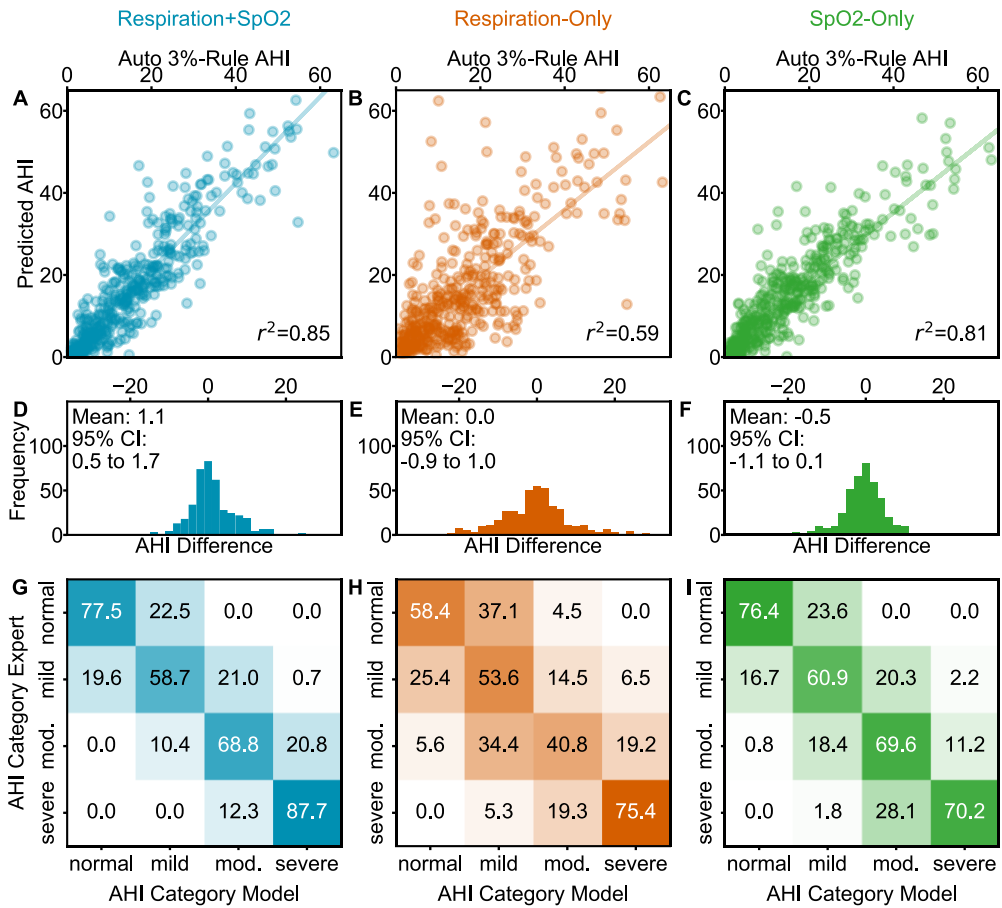

Figure E7. Apnea Hypopnea Index (AHI)-based model performance evaluation (based on auto labels for the 3% hypopnea rule) for models trained on features from respiration and $SpO_2$ (left column), respiration only (middle column), and SpO2-only (right column). Results are obtained from left-out test data in a 10-fold cross-validation fashion; n=409 polysomnographic recordings including wearable respiration. Panels A-C: Scatterplots of predicted (model-based) and expert-labelled AHI per recording. Panels D-F: Difference of predicted and expert-labelled AHI, including 95% confidence interval (CI). Both model results show a unimodal error distribution. Panels G-I: Confusion matrix (in %) with AHI categorization (Normal: 0-5, Mild: 5-15, Moderate: 15-30, Severe: > 30), where rows: expert-labelled AHI category, columns: predicted AHI category. Accuracies for categorizations: 70% (Respiration+$SpO_2$), 54% (Respiration-Only), and 68% (SpO2-Only).

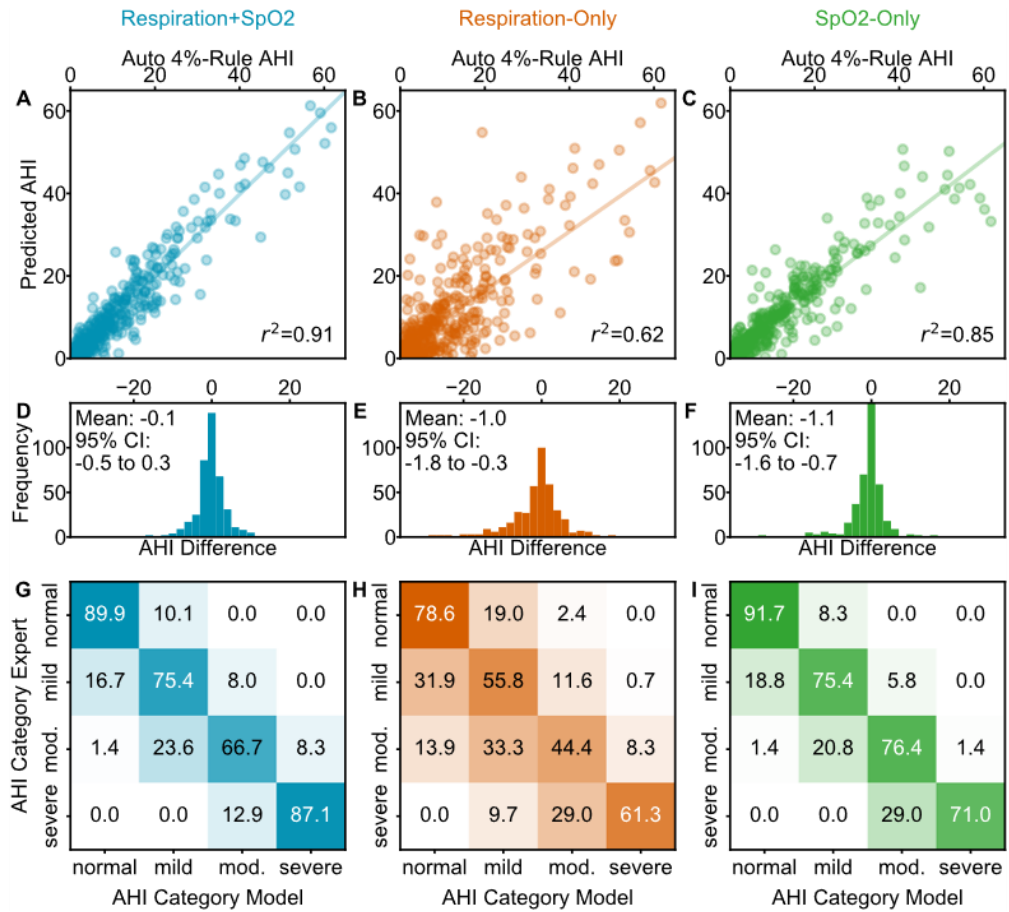

Figure E8. Apnea Hypopnea Index (AHI)-based model performance evaluation (based on auto labels for the 4% hypopnea rule) for models trained on features from respiration and $SpO_2$ (left column), respiration only (middle column), and SpO2-only (right column). Results are obtained from left-out test data in a 10-fold cross-validation fashion; n=409 polysomnographic recordings including wearable respiration. Panels A-C: Scatterplots of predicted (model-based) and expert-labelled AHI per recording. Panels D-F: Difference of predicted and expert-labelled AHI, including 95% confidence interval (CI). Both model results show a unimodal error distribution. Panels **G-I**: Confusion matrix (in %) with AHI categorization (Normal: 0-5, Mild: 5-15, Moderate: 15-30, Severe: > 30), where rows: expert-labelled AHI category, columns: predicted AHI category. Accuracies for categorizations: 81% (Respiration+$SpO_2$), 64% (Respiration-Only), and 82% (SpO2-Only).

TABLE E4 – CORRELATION AHI-SpO$_2$ BURDEN

| AHI ASSESSMENT METHOD | PEARSON CORRELATION |
|---|---|
| AHI Expert Labels 4 | 0.53 |
| AHI Auto-Labels 4 | 0.52 |
| AHI Auto-Labels 3 | 0.46 |
| AHI Resp+SpO$_2$ (EL4) | 0.52 |
| AHI Resp+SpO$_2$ (AL4) | 0.51 |
| AHI Resp+SpO$_2$ (AL3) | 0.50 |
| AHI Respiration-Only (EL4) | 0.37 |
| AHI Respiration-Only (AL4) | 0.39 |
| AHI Respiration-Only (AL3) | 0.37 |
| AHI SpO$_2$-Only (3% threshold) | 0.53 |
| AHI SpO$_2$-Only (4% threshold) | 0.58 |

TABLE E5 – LITERATURE REVIEW TABLE

| | Paper | Year | Database/Hospital | Recordings/Subjects | Wearable/Portable | $R^2$ | Accuracy (%) | AUC |
|---|---|---|---|---|---|---|---|---|
| SpO$_2$ | xx | 2019 | *Database 1 +Database 2 | 8,444 subjects (*Database 1) 218 subjects (+Database 2) | No | - | 81.6, 80.6, 76.6 (3 different test sets) | 0.822, 0.837, 0.816 (3 different test sets) |
| | xxi | 2010 | Hospital | 129 subjects | No | - | 93.02 | 0.95 |
| | xxii | 2016 | Hospital | 127 subjects | Portable | - | 89.8 | - |
| | xxiii | 2014 | Hospital | 34 subjects | No | 0.976 | 93.8 | - |
| | xxiv | 2013 | Hospital | 320 recordings | No | - | 88.7 (FSFS + LR) 84.5 (GAS + SVM) | - |
| | xxv | 2012 | Hospital | 25 subjects | No | - | 83.26 | - |
| | xxvi | 2006 | Hospital | 187 subjects | No | - | 87.2 | 0.92 |
| | xxvii | 2009 | Hospital | 113 subjects | No | - | 87.61 | 0.925 |
| | xxviii | 2017 | Hospital | 79 subjects | No | - | 93.67 (KNN) 88.61 (LS-SVM) | - |
| | xxix | 2012 | Hospital | 144 subjects | No | - | 87.5 | - |
| | xxx | 2012 | Hospital | 115 subjects | No | - | 93.9 | 0.97 |
| | xxxi | 2012 | ^Database | 8 recordings | No | - | 93.3 | - |
| | xxxii | 2008 | Hospital | 187 subjects | No | - | 89.8 | 0.90 |
| | xxxiii | 2017 | ^Database | 8 recordings | No | - | 97.7 | - |
| | xxxiv | 2010 | Hospital | 148 subjects | No | - | 89.7 | 0.967 |
| | xxxv | 2014 | Hospital | 146 subjects | Portable | - | 84.9 | 0.86 |
| | xxxvi | 2017 | ~Database 1 ^Database 2 | 33 subjects | No | - | 85.26 (~Database) 97.64 (^Database) | - |
| | xxxvii | 2017 | ^Database | 8 recordings | Wearable | - | IHR: 89.0 SpO$_2$: 95.5 IHR-SpO$_2$: 92.1 | IHR: 0.88 SpO$_2$: 0.99 IHR-SpO$_2$: 0.99 |
| | xxxviii | 2020 | Hospital | 25 recordings | Wearable | - | 72.8 ± 4.4 | 0.736 ± 0.037 |
| | xxxix | 2018 | Hospital | 1200 data points | Wearable | - | 91.8 | - |
| Respiration Signals | xl | 2019 | *Database | 2100 subjects | No | - | 82.1±0.6 (abdomen) 82.6±0.4 (thoracic) 83.4±0.3 (EDR) | 54.8±0.4 (abdomen) 55.4±0.6 (thoracic) 52.4±0.4 (EDR) |
| | xli | 2015 | ~Database | 25 recordings | No | - | 73.58±5.20 | - |
| | xlii | 2013 | #Database | 28 recordings | Portable | HI: 0.953 AI: 0.976 AHI: 0.984 | HI: 94.3 AI: 96.6 AHI: 96.6 | - |
| | xliii | 2015 | Hospital | 104 signals | No | - | 95 | 0.9837±0.0168 |
| | xliv | 2019 | Hospital | 3 subjects | No | - | 95.8 | - |
| | xlv | 2013 | #Database | 14 recordings | No | - | 95.1 | - |
| | xlvi | 2010 | Hospital | 41 subjects | No | - | - | 0.881 |
| | xlvii | 2012 | Hospital | 148 subjects | No | - | 82.43 | 0.903 |
| | xlviii | 2013 | *Database | 100 recordings | No | - | 72.3 | - |
| | xlix | 2016 | ^Database | 8 recordings | No | - | 98.43 (AdaBoost) 98.68 (ANFIS) | - |
| | l | 2015 | ^Database | 8 recordings | No | - | 98.68 | - |
| | li | 2016 | Hospital | 6 patients | No | - | 87.6 | - |
| | lii | 2017 | %Database | 100 subjects | No | - | 74.70±1.43 | - |
| | liii | 2013 | ^Database | 70 recordings | No | - | 91.2 | 0.9604 |
| | liv | 2011 | Hospital | 12 subjects | No | - | 89.26 | - |
| | lv | 2018 | ~Database | 25 recordings | No | - | 79.61 | - |
| | lvi | 2015 | *Database | 100 recordings | No | - | 95 | - |
| Respiration and SpO$_2$ | lvii | 2020 | Hospital | 239 recordings | Portable | - | AHI > 5: 94.8 AHI > 15: 90.6 AHI > 30: 95.8 | AHI >5: 0.97 AHI > 15: 0.96 AHI > 30: 0.98 |
| | lviii | 2009 | Hospital | 21 subjects | No | - | 80 | - |
| | lix | 2018 | Hospital | 63 subjects | Portable | - | 84.13 | - |

*Database: Sleep Heart Health Study database     #Database: MIT-BIH database

+Database: UZ Leuven database     %Database: MESA sleep study database

Table E6. Event-Based Performance Evaluation for All Sensor-Input and Hypopnea Rules Combinations

Abbreviations: AL3: Automatic Label 3% Hypopnea Rule, AL4: Automatic Label 4% Hypopnea Rule, EL4: Expert Label 4% Hypopnea Rule, TP: True Positives, FP: False Positives, FN: False Negatives, TN: True Negatives, SE: Sensitivity, OA: Obstructive Apnea, CA: Central Apnea, MA: Mixed Apnea, HY: Hypopnea

| Model | Sleep (h) | Events | TP | FP | FN | TN | SE OA | SE CA | SE MA | SE HY |
|---|---|---|---|---|---|---|---|---|---|---|
| Respiration-Only AL3 | 2522 | 27592 | 16175 | 24020 | 11417 | 529,560 | 0.6 | 0.78 | 0.58 | 0.41 |
| Respiration-Only AL4 | 2522 | 27592 | 15479 | 19479 | 12113 | 531,340 | 0.56 | 0.78 | 0.57 | 0.37 |
| Respiration-Only EL4 | 2522 | 27592 | 15334 | 28555 | 12258 | 524,442 | 0.55 | 0.75 | 0.57 | 0.38 |
| Respiration+SpO$_2$ AL3 | 2522 | 27592 | 16323 | 11037 | 11269 | 535,423 | 0.62 | 0.67 | 0.51 | 0.51 |
| Respiration+SpO$_2$ AL4 | 2522 | 27592 | 16736 | 9018 | 10856 | 524,870 | 0.63 | 0.67 | 0.50 | 0.54 |
| Respiration+SpO$_2$ EL4 | 2522 | 27592 | 18411 | 14458 | 9181 | 517,269 | 0.68 | 0.67 | 0.54 | 0.66 |
| Respiration+SpO$_2$ (Robust) AL3 | 2522 | 27592 | 17147 | 14056 | 10445 | 528,875 | 0.65 | 0.71 | 0.54 | 0.53 |
| Respiration+SpO$_2$ (Robust) AL4 | 2522 | 27592 | 17177 | 11527 | 10415 | 520,669 | 0.65 | 0.70 | 0.54 | 0.54 |
| Respiration+SpO$_2$ (Robust) EL4 | 2522 | 27592 | 18647 | 18460 | 8945 | 518,738 | 0.70 | 0.71 | 0.56 | 0.64 |
| SpO$_2$-Only 3% | 2522 | 27592 | 24984 | 17862 | 2608 | 448,211 | 0.92 | 0.83 | 0.77 | 0.97 |

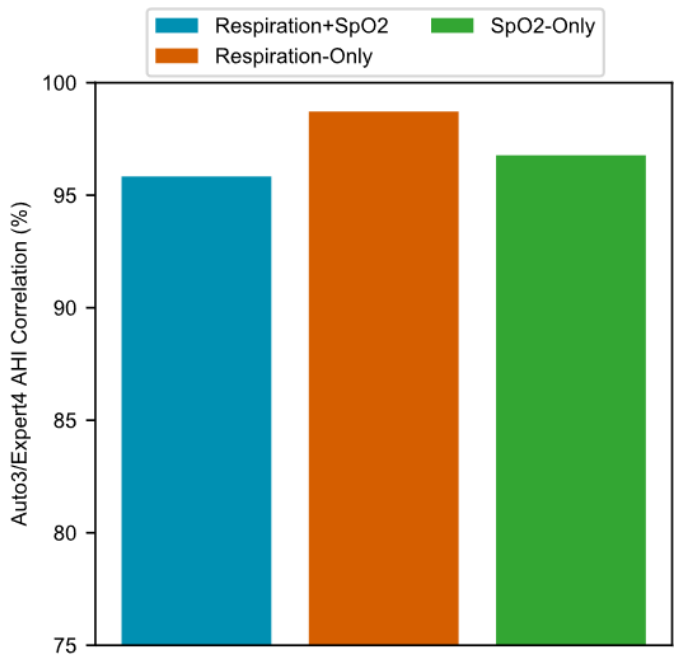
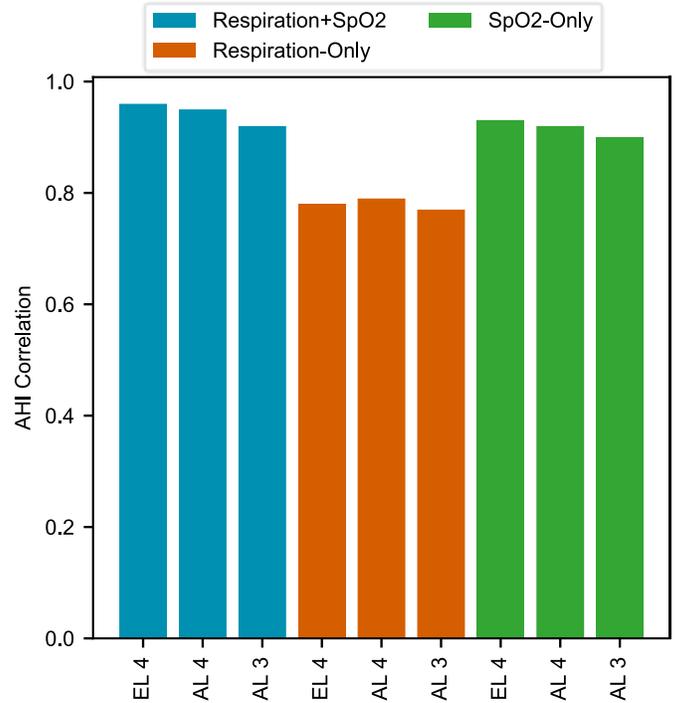

Figure E9. AHI correlation for the different hypopnea labels and different input signals.

Figure E10. Change of AHI correlation between using the expert 4% hypopnea labels and the auto 3% hypopnea labels (in%). Relative to the 4% results, the Respiration-Only achieves largest performance (or the drop in performance is lowest).

References Apendix

iv Richard Bellman, Dynamic Programming (Princeton, NJ, USA: Princeton University Press, 2010).

v Trevor Hastie, Robert Tibshirani, and Jerome Friedman, The Elements of Statistical Learning: Data Mining, Inference, and Prediction, Second Edition, 2nd ed., Springer Series in Statistics (New York: Springer-Verlag, 2009).

vi "Hierarchical Grouping to Optimize an Objective Function: Journal of the American Statistical Association: Vol 58, No 301," accessed April 6, 2020, https://www.tandfonline.com/doi/abs/10.1080/01621459.1963.10500845.

vii Leo Breiman, "Random Forests," Machine Learning 45, no. 1 (October 1, 2001): 5–32, https://doi.org/10.1023/A:1010933404324.

viii Andrew P. Bradley, "The Use of the Area under the ROC Curve in the Evaluation of Machine Learning Algorithms" (Elsevier Science Inc., July 1, 1997), https://doi.org/10.1016/S0031-3203(96)00142-2.

ix "Information Retrieval, 2nd Ed. C.J. Van Rijsbergen. London: Butterworths; 1979: 208 Pp. Price: $32.50 - Blair - 1979 - Journal of the American Society for Information Science - Wiley Online Library," accessed April 6, 2020, https://asistdl.onlinelibrary.wiley.com/doi/abs/10.1002/asi.4630300621.

x Hastie, Tibshirani, and Friedman, The Elements of Statistical Learning.

xi Jesse Davis and Mark Goadrich, "The Relationship between Precision-Recall and ROC Curves," in Proceedings of the 23rd International Conference on Machine Learning - ICML '06 (the 23rd international conference, Pittsburgh, Pennsylvania: ACM Press, 2006), 233–40, https://doi.org/10.1145/1143844.1143874.

xii Berry et al., "Rules for Scoring Respiratory Events in Sleep."

xiii Guido van Rossum and Fred L. Drake, The Python Language Reference Manual (Network Theory Ltd., 2011).

xiv Stefan van der Walt, S. Chris Colbert, and Gael Varoquaux, "The NumPy Array: A Structure for Efficient Numerical Computation," Computing in Science Engineering 13, no. 2 (March 2011): 22–30, https://doi.org/10.1109/MCSE.2011.37.

xv "pandas: a Foundational Python Library for Data Analysis and Statistics | R (Programming Language) | Database Index," Scribd, accessed April 6, 2020, https://www.scribd.com/document/71048089/pandas-a-Foundational-Python-Library-for-Data-Analysis-and-Statistics.

xvi Travis E. Oliphant, "Python for Scientific Computing," Computing in Science Engineering 9, no. 3 (May 2007): 10–20, https://doi.org/10.1109/MCSE.2007.58.

xvii John D. Hunter, "Matplotlib: A 2D Graphics Environment," Computing in Science Engineering 9, no. 3 (May 2007): 90–95, https://doi.org/10.1109/MCSE.2007.55.

xviii Fabian Pedregosa et al., "Scikit-Learn: Machine Learning in Python," Journal of Machine Learning Research 12, no. Oct (2011): 2825–2830.

xix Raphael Vallat, "Raphaelvallat/Entropy," April 4, 2020, https://github.com/raphaelvallat/entropy.

xx Margot Deviaene et al., "Feature Selection Algorithm Based on Random Forest Applied to Sleep ApneaDetection," in 2019 41st Annual International Conference of the IEEE Engineering in Medicine and Biology Society (EMBC), 2019, 2580–83, https://doi.org/10.1109/EMBC.2019.8856582.

xxi J. Víctor Marcos et al., "Automated Detection of Obstructive Sleep Apnoea Syndrome from Oxygen Saturation Recordings Using Linear Discriminant Analysis," Medical & Biological Engineering & Computing 48, no. 9 (September 2010): 895–902, https://doi.org/10.1007/s11517-010-0646-6.

xxii D. Álvarez et al., "Automated Analysis of Unattended Portable Oximetry by Means of Bayesian Neural Networks to Assist in the Diagnosis of Sleep Apnea," in 2016 Global Medical Engineering Physics Exchanges/Pan American Health Care Exchanges (GMEPE/PAHCE), 2016, 1–4, https://doi.org/10.1109/GMEPE-PAHCE.2016.7504628.

xxiii Bijoy Laxmi Koley and Debangshu Dey, "On-Line Detection of Apnea/Hypopnea Events Using SpO2 Signal: A Rule-Based Approach Employing Binary Classifier Models," IEEE Journal of Biomedical and Health Informatics 18, no. 1 (January 2014): 231–39, https://doi.org/10.1109/JBHI.2013.2266279.

xxiv Daniel Alvarez et al., "Assessment of Feature Selection and Classification Approaches to Enhance Information from Overnight Oximetry in the Context of ApneaDiagnosis," International Journal of Neural Systems 23, no. 5 (October 2013): 1350020, https://doi.org/10.1142/S0129065713500202.

xxv Baile Xie and Hlaing Minn, "Real-Time Sleep ApneaDetection by Classifier Combination," IEEE Transactions on Information Technology in Biomedicine 16, no. 3 (May 2012): 469–77, https://doi.org/10.1109/TITB.2012.2188299.

xxvi D. Alvarez et al., "Nonlinear Characteristics of Blood Oxygen Saturation from Nocturnal Oximetry for Obstructive Sleep Apnoea Detection," Physiological Measurement 27, no. 4 (April 2006): 399–412, https://doi.org/10.1088/0967-3334/27/4/006.

xxvii J. Víctor Marcos et al., "Assessment of Four Statistical Pattern Recognition Techniques to Assist in Obstructive Sleep Apnoea Diagnosis from Nocturnal Oximetry," Medical Engineering & Physics 31, no. 8 (October 2009): 971–78, https://doi.org/10.1016/j.medengphy.2009.05.010.
33